# Supporting Car-Following Behavior through V2V-Based Beyond-Visual-Range Information Display

Feiqi Gu[a], Zhixiong Wang[a], Zhenyu Wang[b], Dengbo He[a,b,c]

[a] Thrust of Robotics and Autonomous Systems, Systems Hub, The Hong Kong University of Science and Technology (Guangzhou), Guangzhou, China

[b] Thrust of Intelligent Transportation, Systems Hub, The Hong Kong University of Science and Technology (Guangzhou), Guangzhou, China

[c] HKUST Shenzhen-Hong Kong Collaborative Innovation Research Institute, Guangdong, China

*Corresponding author:

Dengbo He, dengbohe@hkust-gz.edu.cn

No. 1 Duxue Road, Nansha District, Guangzhou, China

**ABSTRACT**

Rear-end collisions constituted a large portion of crashes on the road, despite efforts to mitigate rear-end collisions, such as forward collision warnings. The chance of rear-end collisions is closely related to drivers' car-following (CF) behaviors in the traffic flow. Given that drivers may rely on more than the information of the direct lead vehicle (DLV) when making CF decisions, expanding drivers' perceptual range by providing beyond-visual-range (BVR) information based on vehicle-to-vehicle (V2V) communication may enhance CF safety. Thus, four different human-machine interfaces (HMIs) providing various types of BVR information in CF events were designed, including Brake-HMI showing only brake action of indirect lead vehicles (ILV), Dis-HMI and THW-HMI showing the relative distance and time headway between the ILV and DLV, respectively, and Video-HMI showing the live-stream video of ILV from the perspective of DLV. A driving simulator experiment with 40 participants was conducted to evaluate the impact of BVR-based HMI on driving safety in CF events. We found that, in general, BVR information could improve CF safety without overloading drivers and compromising their visual attention allocation strategies, particularly among novice drivers, by enabling quicker brake responses and increasing time headway and time-to-collision in brake events. The Brake-HMI yielded the safest performance in chain brake events, whereas Video-HMI increased attentional demands without observable benefits. This research provides insights into enabling drivers' BVR perception based on V2V communication to enhance driving safety in CF scenarios.

# 1 INTRODUCTION

According to a statistical report published by the National Safety Council in the United States, rear-end collisions accounted for approximately 28.8% of all reported road incidents in 2022 (Council, 2022). Such collisions commonly result in injuries, particularly whiplash-associated disorders, as well as significant economic losses and property damage (Avery & Weekes, 2009; Siegmund et al., 2009). Previous research has tried to reduce the risk of rear-end collisions by providing drivers with forward collision warnings (FCW) (Adell et al., 2011). However, such information is still ad-hoc, i.e., the FCW can only support the driver's responses to leading vehicle braking after the onset of a critical event.

At the same time, the information asymmetry and delays in information transmission among road users can contribute to traffic accidents, including rear-end collisions (Huang et al., 2023) and congestion phenomena such as phantom traffic jams (Treiber et al., 2000). Thus, expanding drivers' range of perception, i.e., by providing beyond-visual-range (BVR) information, may improve car-following (CF) safety. The BVR information has been made possible in the transportation system thanks to vehicle-to-vehicle (V2V) communication technologies (Yogha Bintoro et al., 2024). However, to date, most of the V2V information has been used to optimize driving automation systems (e.g., Samarakoon et al., 2020). Given that human-driven vehicles may still dominate the market in the next decades (Guo et al., 2021) and that rear-end collisions accounted for an even larger portion of collisions among autonomous vehicles or vehicles with driving automation (Liu et al., 2024), it is necessary to explore how to make better use of the V2V technologies to support human drivers before the fully autonomous and connected vehicles saturate the market.

The research from the human-automation interaction domain may provide some insights. For example, He et al. (2021) found that providing traffic scenario information (i.e., showing the location of surrounding road agents) based on vehicle-to-vehicle (V2V) communication can facilitate drivers' earlier actions to potential traffic hazards in vehicles with driving automation. In CF-related research, our previous preliminary work investigated whether providing braking information of indirect lead vehicles (ILV, i.e., the vehicle ahead of the direct lead vehicle) can facilitate safer CF behaviors in chain-braking events. Though only a video-based experiment was conducted with five participants, we still observed a larger safety margin (as indicated by a larger TTC and earlier brake responses) when ILV information was provided (Yan et al., 2023), which provides evidence to support the role of the BVR information in CF events. However, a limited number of participants were involved in the study, and we have only investigated two types of information visualization – the symbolic one and the video-streaming one, which provided limited information regarding how to design the V2V HMI to provide CF-related BVR information.

As such, a new driving simulation experiment was conducted to investigate how BVR information provided through a connected driving environment can be used to support driver behavior in CF events. Specifically, on top of Yan et al. (2023), we considered both the normal CF scenarios without emergent brakes (but with slight speed variations) and the chain brake scenarios. Further, given that the driving task is already mentally demanding (Salmon et al., 2005), to investigate if additional BVR information would further overload the drivers and how to better provide the ILV-related BVR information, we explored different types of BVR information in the experiment, along with different ways of visualization, including brake actions of the ILV, the distance gap between the direct lead vehicle (DLV) and ILV, the collision risk between these vehicles, as materialized by time headway (THW), and the live video stream from the perspective of the DLV. Given that the driving experience was associated with crash rates (da Silva et al., 2012; McCartt et al., 2003; Mullin et al., 2000), drivers with diverse driving experience were also recruited.

## 2  LITERATURE REVIEW

Various HMI designs have been explored to improve driving safety in CF events. For example, Li et al. (2017) found that the V2V-based warning that alerts drivers of potential hazards in their vicinity can reduce the risk of crashing into the hazards ahead. In a driving simulation experiment, Ali et al. (2020) found that drivers maintained a longer headway to the DLV when a tailgate warning was provided. In another on-road study, Adell et al. (2011) found that safe speed and safe distance reminders can also increase headway in CF events. In addition to the basic safety messages (Liu & Khattak, 2016), Zheng et al. (2023) found that the speed information of the DLV can reduce the drivers' response time to the speed change, resulting in an increase in the minimum time-to-collision (TTC) when the LV slowed down. However, all these previous studies only explored the role of the information from the DLV in CF events. Still, they did not consider the BVR information, such as the relative speed and distance between the ILV and DLV.

At the same time, recent advancements in driving automation and vehicle-to-everything communication have enabled the collection of multiple sources of traffic data beyond drivers' visual range. For example, GPS data from vehicles traversing urban networks can provide information on average travel speeds and congestion in local and even global road networks (Gao et al., 2023). The wide adoption of perception sensors in driving automation systems, such as the LiDAR and cameras, has also made BVR information available (Muhammad et al., 2022). For example, the perception module in intelligent vehicles can now dynamically identify vehicles and obstacles well ahead of the traffic (Zou et al., 2023). Such BVR information has been widely used in traffic control. For example, Olaverri-Monreal et al. (2018) promoted a traffic light assistance system which allowed vehicles to receive real-time traffic signal information, thereby enhancing the decision-making process for drivers and ultimately improving traffic efficiency. However, to date, very few endeavours have been made to use the BVR information to support drivers in CF scenarios.

## 3 METHOD

### 3.1 Participants

A total of 40 gender-balanced participants with valid driver's licenses completed the experiment. Participants had an average age of 27 years old (min: 20, max: 44, SD = 5.9). Half of the participants (10 males and 10 females) were classified as experienced drivers, having held their licenses for over 5 years and driven more than 15,000 kilometers in the past year. The remaining participants (10 males and 10 females) were novice drivers who first obtained their licenses within the past year and had driven less than 5,000 kilometers overall. The study received ethical approval from the Human and Artefacts Research Ethics Committee at the Hong Kong University of Science and Technology (Guangzhou) (protocol number: HREP-2024-0026).

### 3.2 Apparatus

The experiment was conducted using a fixed-based driving simulator by Info Tech, China (Figure 1). Driving scenarios were developed using SILAB 7.1 software (WIVW GmbH, Germany) and projected onto three 43-inch displays, each positioned around one meter away from the driver. Each display had a resolution of 1920*1080, providing a horizontal viewing angle of 150° and a vertical viewing angle of 47°. The driving data was continuously logged at a frequency of 60 Hz within the simulation software. Additionally, during the experiment, participants' eye movement data were recorded at 100 Hz using a remote 4-camera tracking system (Smart Eye Pro) running Smart Eye Pro 10.2.

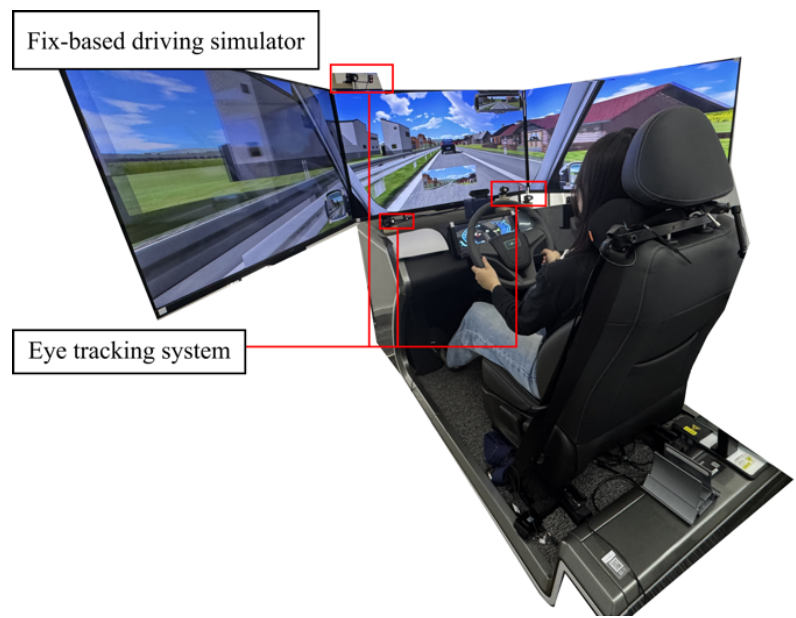

Figure 1. The apparatus used in this study.

### 3.3 HMI Design for BVR Information

In this study, we mainly considered the BVR information regarding the relationships between the DLV and the ILV, as illustrated in Figure 2. In total, four different types of BVR information were presented to the drivers: (1) the braking behavior of the ILV, (2) the bumper-to-bumper distance between the DLV and ILV, (3) the risk of rear-end collision, as represented by the THW between the DLV and the ILV, (4) the live-stream video captured from the perspective of DLV. We visualized the information following Jakob Nielsen's 10 general principles (Nielsen, 1994). All HMI was presented on the bottom of the windshield, simulating the location of the head-up display (HUD). The details of the four design concepts are as follows (see Table 1):

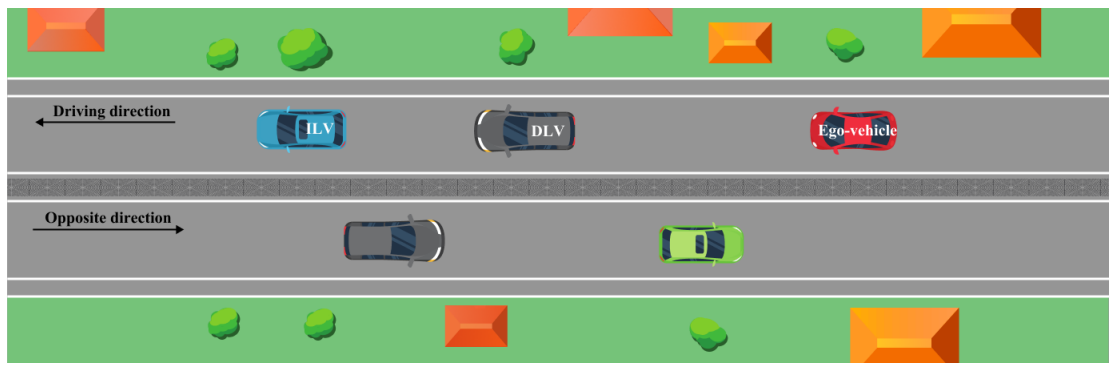

Figure 2. CF scenario in the experiment.

**Brake-HMI**: The brake of the ILV was represented by brake light icons. Once the ILV brakes, the circles become red.

**Dis-HMI**: To represent the bumper-to-bumper distance between the DLV and ILV, two vehicle icons were used, with the space between them representing the actual distance between the DLV and ILV. As the distance increased, the distance between the icons increased, and vice versa. Similarly, the circle on the tail of the ILV icon would represent the ILV braking states.

**THW-HMI**: The risk of a rear-end collision between the DLV and ILV was quantified using THW. When the THW increased, the vehicle icon would become larger and vice versa. Further, a dashed-line circle indicated the 2-second time headway threshold, which has been widely recommended for safe CF behaviors (Lewis-Evans et al., 2010). If the vehicle icon exceeded the circle, it signified a high-risk situation, and vice versa. Similar to the Brake-HMI and the Dis-HMI, when the ILV braked, the two circles on the vehicle icon would become red, informing brake of the ILV.

**Video-HMI**: The live video stream of the ILV shows the real-time video captured from the perspective of the DLV.

Table 1. Four HMI concepts.

| Type of HMI | HMI Visualization | HMIs in the simulated scenario |
|---|---|---|
| **Brake-HMI** | 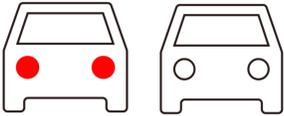 | 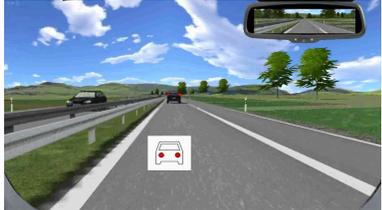 |
| **Dis-HMI** | 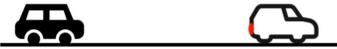 | 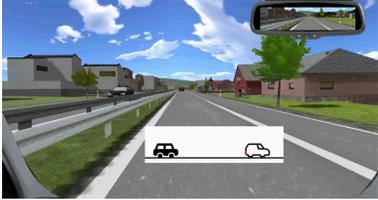 |
| **THW-HMI** | 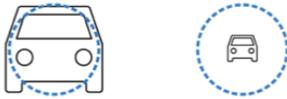 | 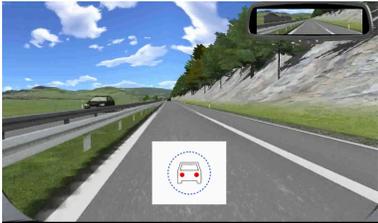 |
| **Video-HMI** | 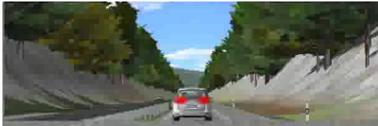 | 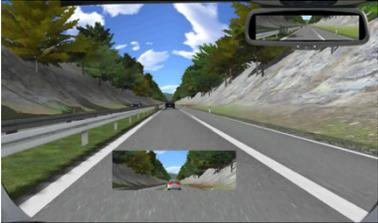 |

## 3.4 Driving Tasks

In the experiment, drivers were required to follow the traffic on a two-lane suburban road on a sunny day, with a speed limit of 60km/h. In the scenario, the ego-vehicle followed a DLV, which further followed an ILV. At the start of the drive, the two lead vehicles drove at the same speed. In each drive, ILV braked six times, with three different rates of deceleration (-1.3m/s$^2$, -5m/s$^2$ and -9.6m/s$^2$) and two initial speeds before brakes (16m/s and 8m/s). In each brake event, the ILV decelerated to 3m/s and kept driving at the speed for 5 to 20 seconds, then accelerated to its initial speed at 2m/s$^2$. Figure 3 illustrates the speed profiles of the ILV and DLV throughout an example drive. The behavior of DLV followed the Intelligent Driver Model (IDM, (Treiber et al., 2000)). During the experiment, participants were instructed to drive safely, adhere to traffic regulations, and avoid rear-end collisions, similar to the scenario when they follow a friend's car to an unknown destination.

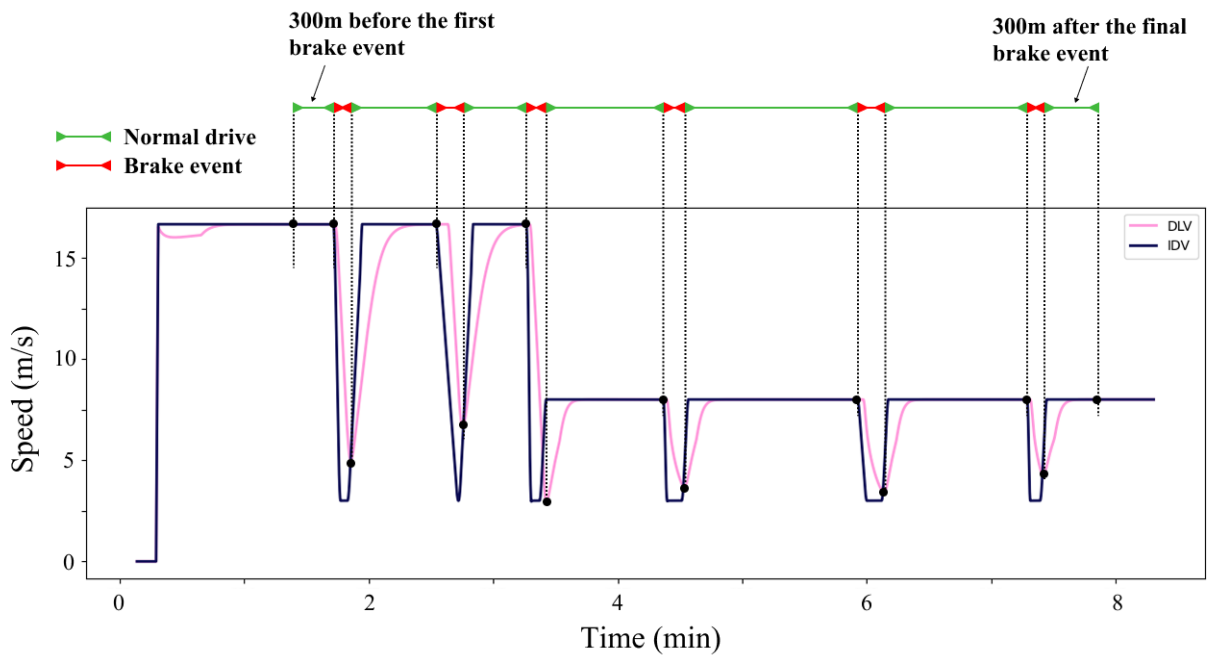

Figure 3. Speed profile of the IDV and DLV in an example drive with periods of interest marked. Note that the order of the speed and the intensity of the braking action was shuffled in the actual experiment.

## 3.5 Experiment Design

A mixed experiment design was used in this study, with the HMI design (i.e., a baseline and four HMIs), the initial speed of ILV (8m/s and 16m/s) and deceleration rate of the ILV (-1.3m/s$^2$, -5m/s$^2$ and -9.6m/s$^2$) as within-subject factors, and driving experience (novice vs. experienced drivers) as a between-subject factor. The order of the HMI design was counterbalanced across the 40 participants in a Latin Square design. Each participant completed five drives, four with the HMI designs and one baseline drive. The order of speed and deceleration rate was shuffled across the five drives, leading to five distinct drives in terms of the speed profile of the ILV to reduce the learning effect. Each participant experienced the same five drives with the same order of speed profiles but with different HMI designs. We collected data from 200 drives (40 participants * 5 HMI designs) and 1200 braking events in total.

## 3.6 Procedures

As shown in Figure 4, upon arrival, participants' eligibility was verified based on years of licensure and driving mileage over the past year, and written informed consent was obtained. Then, they completed a pre-experiment questionnaire, collecting their demographic information. Participants then underwent training with the driving simulator. They first received verbal instructions, followed by a practice drive without the HUD to familiarize themselves with the simulator environment. The elements of the driving scenario in the training session (including the road type and the traffic conditions) were the same as those in the experimental drives, but without ILV or DLV. After finishing the training drive, the experimenter conducted the eye tracker calibration.

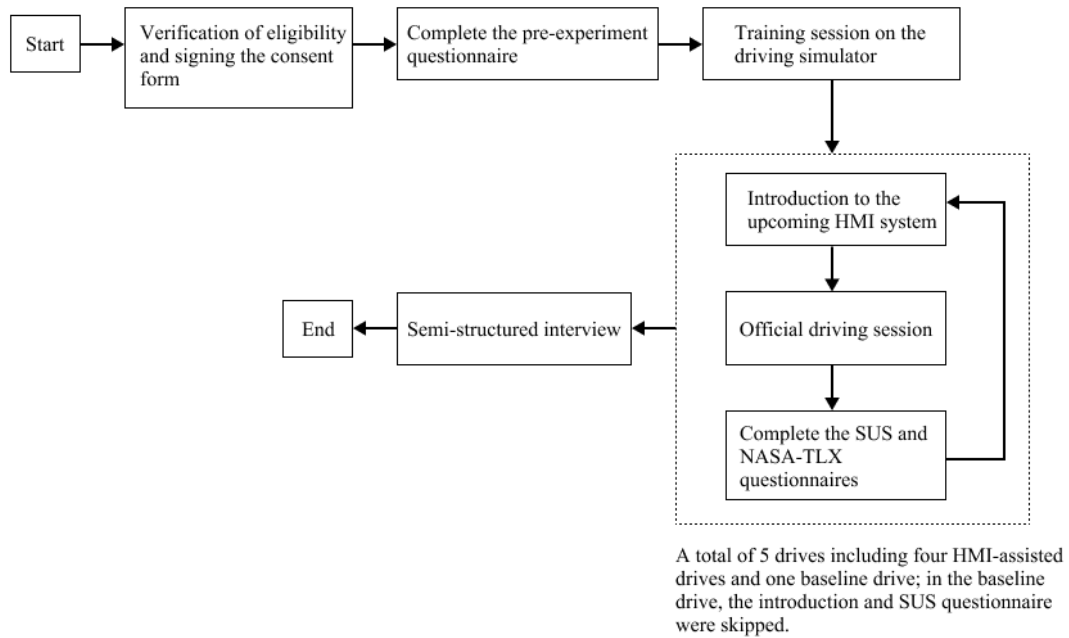

Figure 4. Flow diagram of experiment procedure.

Then, the formal experiment started, which included five experimental drives. Before each drive with the designed HMIs, the HMI design was explained to the participants verbally with a pre-recorded 30-second video showing the animation of the HMI. After the explanation, the experimenter asked the participants to explain the HMI design back to the experimenter to make sure they indeed understood HMI. After each experimental drive, a post-experiment questionnaire was administered, measuring participants' workload in the previous drive (measured by NASA-Task Load Index (Hart & Staveland, 1988)) and perceived usability of the HMI (measured by System Usability Scale (SUS) (Brooke, 1996)). Finally, at the end of the experiment, we asked an open question seeking any comments regarding the HMI designs in the experiment.

## 4 DATA ANALYSIS

We first focused on driving performance data to evaluate the effect of HMI on drivers' CF behaviors. Further, we evaluated drivers' eye-tracking data given that HMIs may affect driver's attention allocation strategies. Finally, to evaluate the usability of the HMI designs, we compared drivers' workload and perceived usability of the HMI designs when using different HMIs.

At the same time, as shown in Figure 3, to comprehensively evaluate the impact of HMI on drivers' behaviors, we focused on two periods of interest in the experiment, i.e., the brake events (from ILV brake onset to the DLV stopping braking) and the normal drive (drive sections outside of brake sections). All driving performance and eye-tracking metrics were extracted from one or both sections, as detailed in Table 2.

### 4.1 Driving Performance Data

In this study, as shown in Table 2, we adopted four driving-performance-related metrics. It should be noted that, for the response time. A total of 14 response times (RTs) smaller than 0.1 s or larger than 5 s were discarded from the data analysis, as such RTs were considered irrelevant to the stimuli (i.e., the brake of the ILV) (Zheng et al., 2023).

Table 2. Driving performance metrics, their definitions and data extraction sections.

| Metric (Abbreviation) | Definitions (Unit) | Periods of interest |
| --- | --- | --- |
| Response time (RT) | The brake time difference between ILV and the ego-vehicle (ms). | Brake event |
| Minimum time headway (MinTHW) | The minimum elapsed time during braking events between the front of the DLV and the front of the ego-vehicle passing the same fixed point on the roadway (s). | Brake event |
| Minimum time-to-collision (MinTTC) | The time required for ego-vehicle to collide with DLV if they continue at their current speeds and on the same path in the current lane (s) | Brake event |
| Average time headway (MeanTHW) | The mean elapsed time during braking events between the front of the DLV and the front of the ego-vehicle passing the same fixed point on the roadway (s) | Normal drive |

**4.2 Eye-Tracking Metrics**

The eye-tracking data was recorded using Smart Eye Pro 10.2. In this study, eye movement data was analyzed during the normal drive in order to understand the impact of the additional information provided by the HMI on drivers' visual attention allocation. Inspired by Ezzati Amini et al. (2023), as shown in Figure 5, we defined five areas of interest (AOI) for the analysis of eye-tracking metrics, including road ahead, roadside, rear-view mirror, dashboard and HMI.

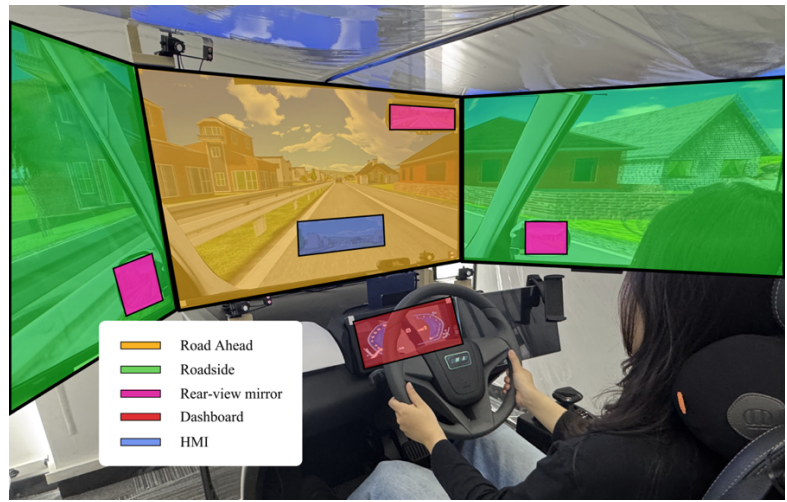

Figure 5. Demonstration of AOIs.

For the eye-tracking measures, we followed the ISO 15007–1:2013(E) standard to process and extract the corresponding metrics. Specifically, each glance was defined as the moment the gaze shifted toward an AOI until it started to move away from it. Following Crundall and Underwood (2011), glances smaller than 100ms were excluded from analysis. Based on the data collected by the eye tracker, we calculated the percent of time looking at the specific AOI during normal drive sections to understand the distribution of visual attention across different AOIs when different HMIs were provided. The mean glance durations on the HMI areas were also analyzed in order to evaluate the potential distracting effect of HMIs, as long off-road glances can increase the crash risks (Horrey & Wickens, 2007). Besides, we also extracted gaze-related metrics, including the vertical and horizontal gaze dispersion, to compare the ranges of visual attention when different HMIs were provided (Huang et al., 2024; Pillai et al., 2022).

## 4.3 Subjective Metrics

We followed the standard approach (Hart & Staveland, 1988) to calculate the weighted overall workload of drivers throughout each drive, leading to 200 data points (40 participants * 5 drives). As for the perceived usability of the HMIs, we followed the standard approach (Brooke, 1996) to calculate the SUS score for each HMI, again, leading to 160 data points.

## 4.4 Independent Variables and Statistical Models.

Two independent variables and their interaction effect were included in the model, i.e., driving experience and HMI design. Different models were built to investigate whether the HMIs were effective in different traffic conditions. Mixed-effects linear models were fitted using the lme4 package in R. Post-hoc comparisons were conducted for significant main effects or interaction effects ($p < 0.05$), using the Turkey adjustment method to control for multiple comparisons. Logarithmic transformation was applied to the dependent variables when the residuals of the model violated the normality assumption (including minTHW, minTTC, meanTHW, percentage of glance duration, and mean glance duration).

## 5 RESULTS

### 5.1 Driving Performance Metrics

Table 3 summarizes the results regarding the influence of HMIs on the driving performance metrics.

Table 3. Results of driving performance metrics.

| Dependent variables | Independent variables | F-value | *p*-value |
|---|---|---|---|
| RT (brake event) | **HMI** | **F (4, 752) = 10.97** | **<.0001** |
| | Driving experience | F (1, 37) = 0.09 | .7 |
| | HMI* Driving experience | F (4, 752) = 0.71 | .6 |
| MinTHW (brake event) | **HMI** | **F (4, 572) = 0.74** | **.03** |
| | **Driving experience** | **F (1, 18) = 4.60** | **.0005** |
| | **HMI* Driving experience** | **F (4, 572) = 4.00** | **.04** |
| MinTTC (brake event) | **HMI** | **F (4, 1125) = 2.76** | **.03** |
| | **Driving experience** | **F (1, 38) = 6.63** | **.01** |

| | HMI* Driving experience | F (4, 1125) = 0.38 | .8 |
| --- | --- | --- | --- |
| MeanTHW (normal drive) | HMI | F (4, 90) = 0.14 | .9 |
| | **Driving experience** | **F (1, 90) = 14.97** | **.0002** |
| | HMI* Driving experience | F (4, 90) = 0.28 | .9 |

*Note:* the significant main or interaction effects are bolded in the table.

**RT:** As shown in Table 3 and Figure 6, HMIs were found to have significant effect on RT. Compared with baseline, Brake-HMI reduced RT, with an estimated difference (Δ) of -0.65 sec, 95% confidence interval (95%CI) of [-0.99, -0.31], t (793) = -5.25, *p* < .0001. At the same time, THW-HMI led to reduced RT compared to baseline (Δ = - 0.66 sec, 95% CI: [-1.00, -0.33], t (789) = -5.38, *p* < .0001), Dis-HMI (Δ = - 0.42 sec, 95%CI: [-0.75, -0.08], t (787) = -3.42, *p* = 0.006) and Video-HMI (Δ = -0.34 sec, 95%CI: [-0.67, -0.004], t (787) = -2.77, *p* = .046). Finally, RT of Brake-HMI was significantly smaller than that of Dis-HMI (Δ = -0.41 sec, 95%CI: [-0.74, -0.07], t (788) = -3.31, p = .009).

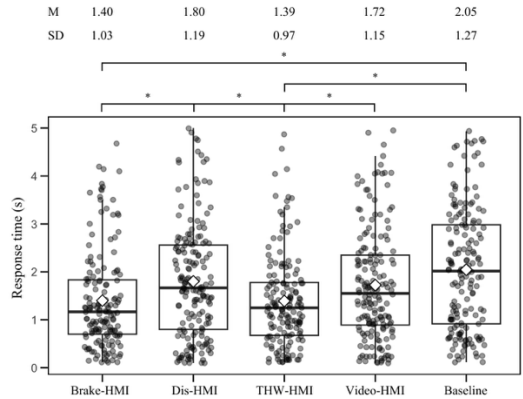

Figure 6. Post hoc comparisons of the effects of HMIs on RT. In this figure and the following figures, significant post-hoc comparisons ($p < 0.05$) are marked with "*"; the boxplot represents the 1st quantile, median, and 3rd quantile; the white squares are the mean of the group.

**MinTHW:** As shown in Figure 7, in brake events, we observed a significant interaction effect between the HMI and the driving experience. Specifically, for novice drivers, the Brake-HMI, Dis-HMI and THW-HMI all led to larger minTHW than the baseline (Brake-HMI vs. baseline: $\Delta \ln = 0.67$, 95%CI: [0.14, 1.20], t (328) = 3.99, $p$ = .003; Dis-HMI vs. baseline: $\Delta \ln$=0.57, 95%CI: [0.08, 1.07], t (425) = 3.65 $p$ = .01; THW-HMI vs. baseline: $\Delta \ln = 0.55$, 95%CI: [0.09, 1.00], t (653) = 3.80, $p$ = .006). At the same time, in baseline condition, experienced drivers exhibited larger minTHW than novice drivers ($\Delta \ln = 0.84$, 95%CI: [0.28, 1.39], t (215) = 4.80, $p < .0001$).

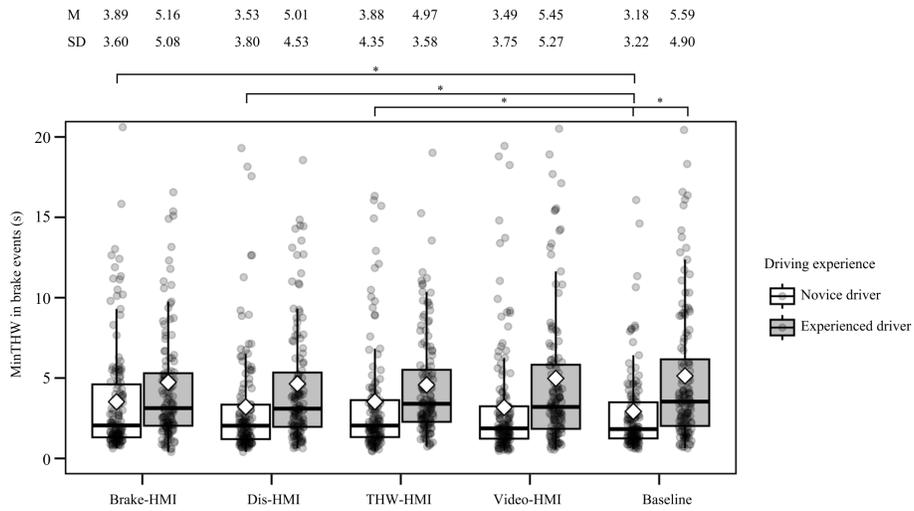

Figure 7. Post hoc comparison of the interaction effect between HMI and driving experience on minTHW in brake events.

**MinTTC:** In the brake events, as shown in Figure 8, Brake-HMI led to a larger minTTC than baseline (Δln = 0.19, 95%CI: [0.009, 0.37], t (1126) = 2.86, *p* = .03). At the same time, novice drivers had lower minTTC than experienced drivers (Δln = -0.38, 95%CI: [-0.68, -0.08], t (38) = -2.58, p = .01)

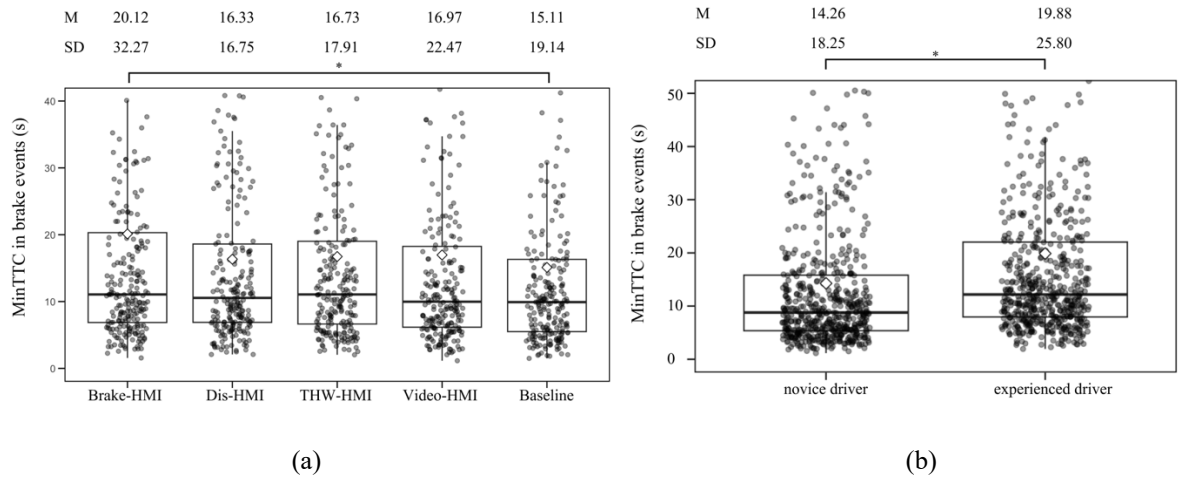

(a)                  (b)

Figure 8. Post hoc comparison of the minTTC: a) the effect of HMI in brake events; b) the effect of driving experience in brake events.

**MeanTHW:** In normal drive sections, only a significant driving experience effect was observed, with novice drivers exhibiting a smaller meanTHW than experienced drivers (Figure 9, $\Delta \ln = -0.36$, 95%CI: [-0.67, -0.05], $t(43) = -2.34$, $p = .02$).

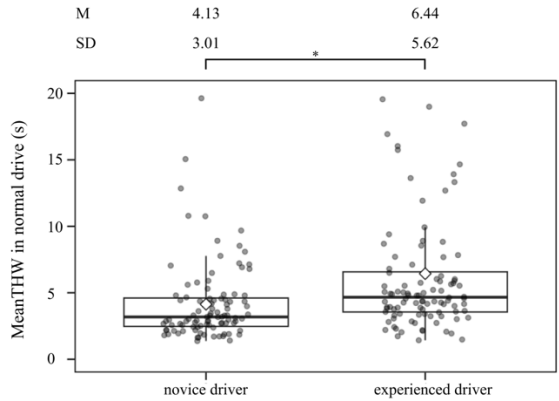

Figure 9. Post hoc comparison of the effect of driving experience on meanTHW in normal drive.

## 5.2 Eye-Tracking Metrics during Normal Drive Sections

Table 4 summarizes the results for the influence of HMIs on drivers' visual attention allocation during normal drive sections.

Table 4. Results of eye-tracking metrics.

| Dependent variables | Independent variables | F-value | p-value |
|---|---|---|---|
| Percent of time looking at HMI area | **HMI** | **$F(4, 110) = 43$** | **<.0001** |
| | Driving experience | $F(1, 36) = 0.78$ | .4 |
| | HMI* Driving experience | $F(4, 110) = 0.52$ | .7 |
| | HMI | $F(4, 180) = 0.91$ | .9 |

| Dependent variables | Independent variables | F-value | p-value |
| --- | --- | --- | --- |
| Percent of time looking at dashboard | Driving experience | $F(1, 25) = 1.25$ | .3 |
| | HMI* Driving experience | $F(4, 80) = 0.9$ | .9 |
| Percent of time looking at road ahead | **HMI** | **$F(4, 143) = 4.72$** | **.001** |
| | Driving experience | $F(1, 37) = 0.02$ | .8 |
| | HMI* Driving experience | $F(4, 143) = 0.87$ | .4 |
| Percent of time looking at roadside | HMI | $F(4, 86) = 0.47$ | .8 |
| | Driving experience | $F(1, 28) = 1.20$ | .3 |
| | HMI* Driving experience | $F(4, 86) = 1.85$ | .1 |
| Percent of time looking at rear-view mirrors | HMI | $F(4, 80) = 1.56$ | .2 |
| | Driving experience | $F(1, 28) = 1.37$ | .2 |
| | HMI* Driving experience | $F(4, 80) = 0.84$ | .5 |
| Mean duration of looking at HMI area | **HMI** | **$F(4, 113) = 7.75$** | **<.0001** |
| | Driving experience | $F(1, 39) = 0.02$ | .9 |
| | HMI* Driving experience | $F(4, 113) = 1.90$ | .1 |
| SD of horizontal gaze – | HMI | $F(4, 143) = 1.65$ | .2 |
| | Driving experience | $F(1, 37) = 0.66$ | .4 |
| | HMI* Driving experience | $F(4, 143) = 0.71$ | .6 |
| SD of vertical gaze | HMI | $F(4, 142) = 1.91$ | .1 |
| | Driving experience | $F(1, 37) = 3.79$ | .06 |
| | HMI* Driving experience | $F(4, 142) = 0.81$ | .5 |

**Percent of Time Looking at Different AOIs:** The HMI type significantly influenced the percent of time looking at HMIs. Specifically, as shown in Figure 10, Figure 11 and Table 5, with Dis-HMI, THW-HMI, and Video-HMI, drivers spent a significantly larger percent of time looking at the HMI area than the baseline, with the Video-HMI attracting most of the visual attention. No significant difference in the time looking at HMIs between Brake-HMI and baseline.

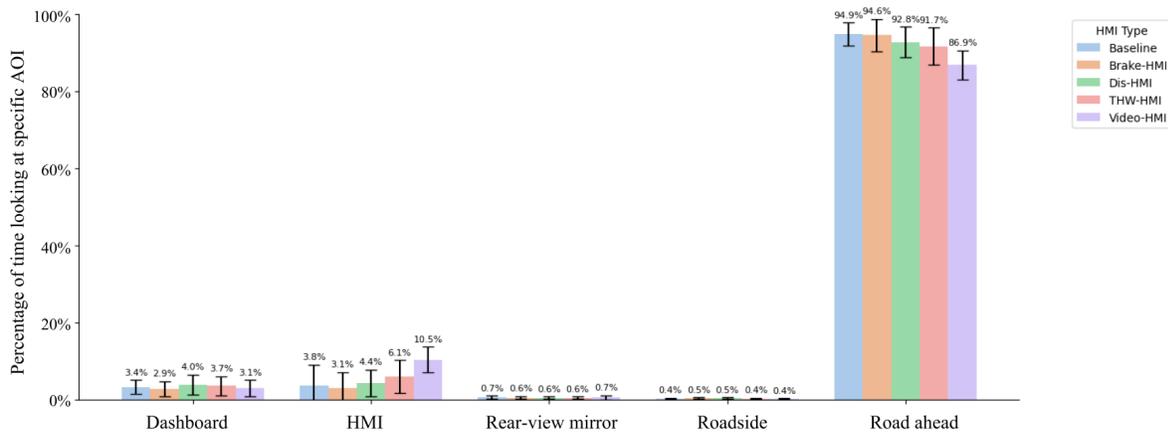

Figure 10. Percent of time looking at specific AOIs under different HMI conditions.

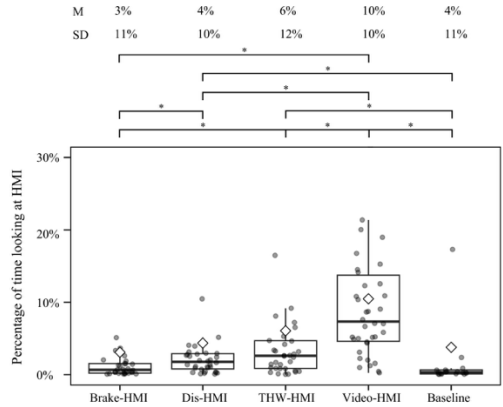

Figure 11. Post hoc comparison of the effect of HMI on percent of time looking at HMI area.

Table 5. Significant pairwise comparisons for the percent of time looking at the HMI area.

| HMI condition | Δln (95%CI) | t-value | p-value |
| --- | --- | --- | --- |
| Brake-HMI vs. Dis-HMI | -0.86 (-1.53, -0.19) | t (109) = -3.54 | .005 |
| Brake-HMI vs. THW-HMI | -1.19 (-1.86, -0.51) | t (110) = -4.89 | <.0001 |
| Brake-HMI vs. Video-HMI | -2.48 (-3.13, -1.82) | t (110) = -10.51 | <.0001 |
| Dis-HMI vs. Video-HMI | -1.62 (-2.25, -0.99) | t (110) = -7.11 | <.0001 |
| Dis-HMI vs. Baseline | 1.49 (0.69, 2.28) | t (112) = 5.18 | <.0001 |
| THW-HMI vs. Video-HMI | -1.29 (-1.92, -0.66) | t (109) = -5.70 | <.0001 |
| THW-HMI vs. Baseline | 1.82 (1.02, 2.61) | t (111) = 6.35 | <.0001 |
| Video-HMI vs. Baseline | 3.11 (2.33, 3.89) | t (111) = 11.09 | <.0001 |

As shown in Figure 10 and Figure 12, Video-HMI led to significantly lower percent of time looking at the road ahead area compared to Brake-HMI, Dis-HMI and baseline (Video-HMI vs. Brake-HMI: Δln = 0.07, 95% CI: [0.009, 0.13], t (142) = 3.16, $p$ = .0002; Video-HMI vs. Dis-HMI: Δln = 0.07, 95% CI: [0.003, 0.13], t (143) = 2.89, $p$ = .04; Video-HMI vs. Baseline: Δln = 0.09, 95% CI: [0.03, 0.15], t (142) = 3.91, $p$ = .001).

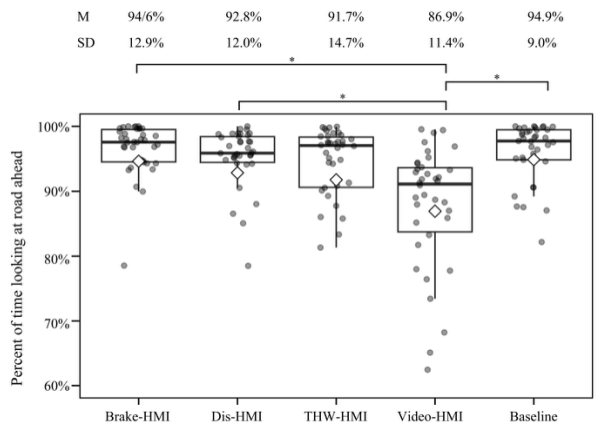

Figure 12. Post hoc comparison regarding the effect of HMI on the percent of time looking at road ahead area.

**Mean Duration Looking at HMI:** As shown in Figure 13 and Table 6, Video-HMI led to significantly longer mean duration looking at the HMI area than other HMIs and baseline. THW-HMI led to a longer mean duration than baseline.

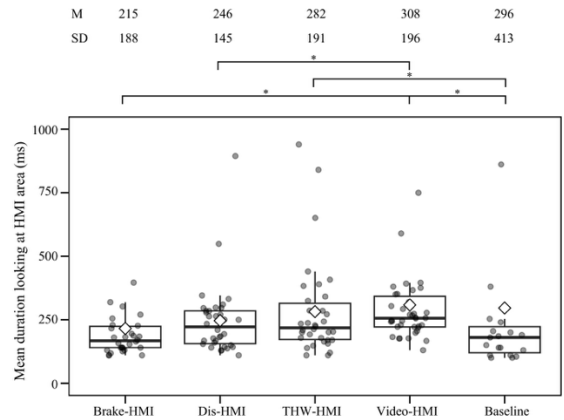

Figure 13. Post hoc comparisons of the effect of HMI on mean duration looking at HMI area.

Table 6. Significant pairwise comparisons for mean duration looking at HMI area.

| HMI condition | Δln (95%CI) | t-value | p-value |
|---|---|---|---|
| Brake-HMI vs. Video-HMI | -0.40 (-0.65, -0.16) | t (112) = -4.53 | .0001 |
| Dis-HMI vs. Video-HMI | -0.24 (-0.48, -0.01) | t (113) = -2.81 | .045 |
| THW-HMI vs. Baseline | 0.32 (0.02, 0.62) | t (115) = 2.96 | .03 |
| Video-HMI vs. Baseline | 0.47 (0.18, 0.77) | t (115) = 4.51 | .0002 |

**SD of Horizontal and Vertical Gaze:** No significant association was observed between HMI and SD of vertical and horizontal eye movements ($p > .05$).

## 5.3 Mental Workload and Perceived Usability

Table 7 summarizes the results of drivers' perceived usability, learnability of the HMIs and the corresponding workload.

Table 7. Statistical analysis results.

| Dependent variables | Independent variables | F-value | *p*-value |
|---|---|---|---|
| Perceived usability | **HMI** | **F (3, 114) = 3.44** | **.02** |
| | Driving experience | F (1, 38) = 0.01 | .9 |
| | HMI* Driving experience | F (3, 114) = 1.01 | .4 |
| Perceived learnability | **HMI** | **F (3, 114) = 4.09** | **.008** |
| | Driving experience | F (1, 38) = 0.68 | .4 |
| | HMI* Driving experience | F (3, 114) = 10.31 | .8 |
| Mental workload | HMI | F (3, 158) = 0.51 | .5 |
| | Driving experience | F (1, 63) = 0.58 | .8 |
| | HMI* Driving experience | F (3, 158) = 0.73 | .4 |

As shown in Table 7, neither HMI type nor driving experience affected mental workload ($p > .05$). At the same time, as shown in Figure 14, Brake-HMI yielded higher perceived usability than Video-HMI ($\Delta = 8.61$, 95%CI: [0.92, 16.31], t (115) = 2.92, $p = .02$) and higher perceived learnability than THW-HMI (Δ = 10.62, 95%CI: [4.40, 16.84], t (114) = 3.38, *p* = .005). No other significant results were observed (*p* > .05).

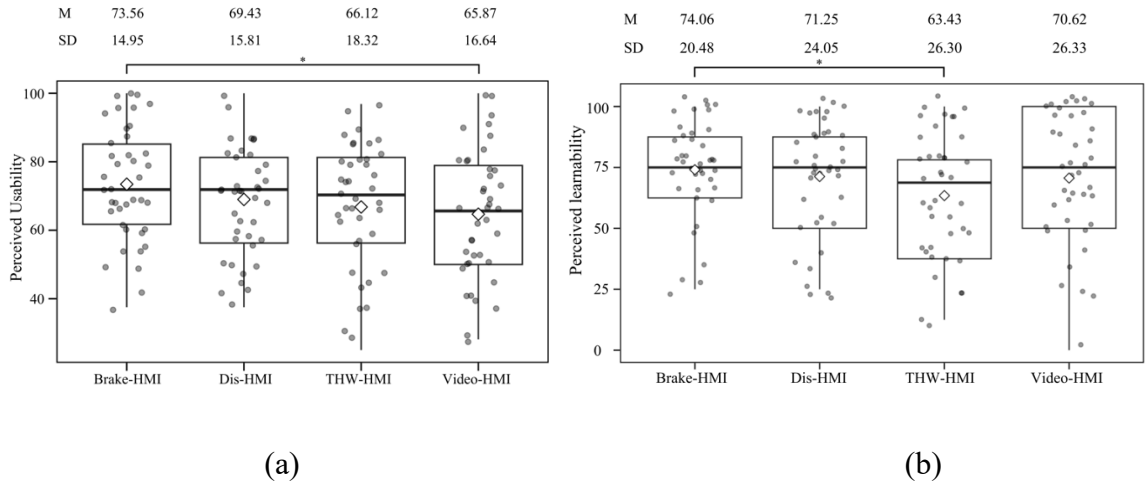

Figure 14. Post hoc comparison of the perceived a) usability and b) learnability of HMIs.

## 6 DISCUSSION

A driving simulator study was conducted to explore the effect of BVR information on drivers' performance in CF events. We evaluated participants' driving behavior, visual behavior, and subjective attitudes with different BVR information.

### 6.1 Effect of BVR Information on Driving Performance

First, the HMI condition did not affect the MeanTHW in the drive sections without braking events, which indicated that drivers' CF strategies may still be primarily based on the relative motion between the ego-vehicle and the DLV in normal driving. However, we found that drivers in general exhibited safer driving behaviors in response to lead vehicle chain braking events when supported by BVR information. Specifically, the Brake-HMI and THW-HMI led to a shorter RT than baseline, possibly by reminding the drivers of the brake of ILV in advance. These results echo the findings in (Yan et al., 2023) who also found ILV braking information can facilitate quicker responses to the brake of lead vehicles, providing further empirical evidence that drivers would consider the surrounding information when following lead vehicles, other than relative distance to the DLV (Han et al., 2022). However, neither Dis-HMI nor Video-HMI yielded shorter RTs. It is possible that both Brake-HMI and THW-HMI contained obvious information about the brake signal of ILVs (the red icons on the top of the vehicle icon), while the ILV brake signals in Dis-HMI and Video-HMI were less obvious in the HMI. Thus, future HMI design may need to explicitly convey BVR information with high salience to support drivers' reactions.

Such a difference in RT also led to improved safety performance in the brake events, especially among novice drivers. Specifically, all HMIs except Video-HMI led to larger MinTHW among novice drivers in the braking events, and the Brake-eHMI further increased the MinTTC in the braking event. The lack of difference regarding the MinTHW between the HMI conditions among the experienced drivers might be attributed to the already good performance of the experienced drivers, given that even in the baseline without any HMIs, experienced drivers have kept a larger MinTHW and MinTTC compared to novice drivers. This has also been supported by the superior performance of experienced drivers in responding to emergent braking events in previous research (Loeb et al., 2015).

Further, we did not observe the effects of Dis-HMI on the RT, but still observed a significant effect of Dis-HMI on MinTHW. It is possible that the Dis-HMI provided drivers with rich and easy-to-perceive information regarding the relative distance and speed between the ILV and DLV. Thus, drivers tended to brake "right on time" instead of braking earlier. This may have also explained why only the Brake-HMI led to shorter MinTTC – drivers tended to brake anyway once they noticed the ILV braking, if they did not have additional information to help them judge the level of emergency in the event (as has been provided in Dis-HMI and THW-HMI). This result also aligns with the relatively high learnability of the brake-HMI – it is straightforward for the drivers to perceive, understand and respond directly. However, it should be noted that such a simple response to ILV braking may not always be good, as it may potentially worsen the stability of the traffic flow and increase the fuel consumption (Horn, 2013) if the braking is unnecessary (e.g., when the ILV is far away from the DLV).

Finally, it should be noted that the Video-HMI failed to exhibit any effects on the safety metrics. It is possible that the drivers had difficulty judging the relative distance between the ILV and DLV based on the visual angle of the DLV alone, especially when the DLV was far away (Flannagan et al., 1997). In contrast, both Dis-HMI and THW-HMI have explicitly visualized the relative distance or time headway between the ILV and DLV. Thus, given that driving is already an attentionally demanding task (Salmon et al., 2005), explicitly visualizing implicit information may reduce drivers' workload and support better driver decisions, especially in emergent events.

**6.2 Effect of BVR Information on Drivers' Attention Allocation**

To better explain the difference in driving performance when different types of HMIs were used, we compared drivers' visual attention allocation when different HMIs were provided. We found that the amount of visual attention attracted by the HMIs was roughly proportional to the amount of information in HMIs. Specifically, echoing the results of driving performance metrics, though not all pairwise comparisons were significant, in general, out of all HMIs, Video-HMI attracted the most visual attention among all HMIs and the Brake-HMI attracted the least amount of visual attention, as indicated by the percent of time spent on the HMI area, the percent of time on the road ahead and the mean glance duration on the HMI areas. This supports our previous assumptions that perceiving the relative distance and motion between the DLV and the ILV based on video alone can be attentionally demanding. Such results also aligned with participants' subjective ratings of the HMIs, with the Video-HMI being rated as having a relatively low level of usability, which also agrees with the conclusions from Yan et al. (2023). These results underscore the importance of balancing informativeness and usability in HMI design. In other words, the Video-HMI contained the richest information among all HMIs (as the ILV braking, time headway and distance can all be perceived from the live video captured from the perspective of the DLV), but providing more information in an HMI may not always be good, especially during an already attention-demanding task such as driving.

Finally, the lack of difference in the perceived workload (NASA-TLX) among all HMIs implies that the additional BVR information provided by these HMIs did not significantly overload the drivers. Further, none of the HMIs yielded different gaze dispersion than baseline. Given that the increased variance of fixation locations, as indicated by broader scanning areas (Robbins & Chapman, 2019), was associated with the increased driving experience and thus lower crash risk (Lehtonen et al., 2014; Underwood et al., 2003), all our HMIs designs may not have compromised driving safety.

**6.3 Limitation**

In this study, we only considered the difference between novice and experienced. However, the driving style may also affect the CF strategies (Motonori et al., 2007) and they may perceive the HMIs differently, which should be carefully evaluated if any of the HMIs are to be deployed in the real world. Second, we only evaluated the HMIs from the safety perspective; future research should evaluate the impact of the HMIs on the traffic flow stability and the environmental impact of the HMIs. Finally, all participants were explicitly informed of the meaning of the HMIs, assuming that the HMIs can be standardized in the future. However, future research should evaluate whether such HMIs can be understood when the drivers were not informed of the meaning of the HMIs, assuming these HMIs are deployed in mass-production vehicles directly.

# 7 CONCLUSIONS

In a driving simulator experiment, we evaluated four types of BVR HMIs in CF scenarios. The results showed that:

- The BVR information can potentially improve CF safety, especially among novice drivers.
- Simple and intuitive Brake-HMI showing the ILV brake actions can significantly facilitate faster brake response time, and increase time headway and TTC in brake events, especially among novice drivers.
- In contrast, complex displays with the richest information, i.e., the Video-HMI providing video streams, increased attentional demands but without yielding obvious performance gain.
- Though BVR HMIs attracted drivers' visual attention, they did not alter drivers' visual attention strategies and did not overload drivers in CF scenarios.

Overall, the BVR information has the potential to enhance driving safety if well-designed to balance informativeness, complexity and intuitiveness. Future research should explore adaptive systems that tailor BVR displays to driver experience and context, supporting safer and more efficient driving in connected vehicle environments.

## ACKNOWLEDGMENT

This work was supported by the National Natural Science Foundation of China (No. 52202425), and in part by the Guangzhou Municipal Science and Technology Project (No. 2023A03J0011), and the Guangdong Provincial Key Lab of Integrated Communication, Sensing and Computation for Ubiquitous Internet of Things (No. 2023B1212010007).